\documentstyle[fleqn,11pt]{article}
\title{
\hskip5.2in\parbox[t]{1.25in}{\large\raggedright NS--USTC/93-50 revi
\\Dec 27, 1993}\\
\vskip1.25cm
Second Order Perturbation Corrections to Baryon Mass Spectrum in Skyrmion
Quantum Mechanics
\thanks{The work was supported by National Theoretical
Physics Project, National
Science Foundation and Science Academy
Foundation in P.R.China.}}\author{
{\large  Mu-Lin Yan and Xin-He Meng}\\
Fundamental Physics Center\\
University of Science \& Technology of China\\Hefei Anhui 230026, P.R.China}
\date{Revised, Dec 27, 1993}

\begin{document}
\baselineskip0.4in
\maketitle
\begin{abstract}
{ The corrections to the Gell-Mann-Okubo relations of baryon-masses are
presented in the SU(3) Skyrme model. These corrections are calculated
up to the second order in flavor breaking at the skyrmion quantum mechanics.
The results are compatible with the experimental data.
They could be regarded
as a signal of existence of some obscure SU(3)-multiplets: 27-let(with
spin 1/2 or 3/2),10$^*$-let(with spin 1/2) and 35-let(with spin 3/2).
}
\end{abstract}
\newpage

   1. The Gell-Mann-Okubo relations (GOR) of baryon-masses$^{[1]}$ were
originally formulated in terms of a perturbative treatment of flavor-
breaking in SU(3) group theory. In the history, the success of the GOR
(or SU(3) theory) led to the quark conception and deep understandings
for strong interactions. In recent years, it is attented
that the high order effects in flavor breaking in some effective
theories of QCD could reach new mass formulas and improve the GOR
$^{[2][3]}$. According to GOR, the baryon-octet and -decuplet
mass-relations can be written as
\begin{equation}
2(M_{N}+M_{\Xi})=3M_{\Lambda}+M_{\Sigma}+\delta m_{8},\\
\end{equation}
\begin{equation}
M_{\Delta}-M_{\Sigma^{*}}=M_{\Sigma^{*}}-M_{\Xi^{*}}+\delta
 m_{10}^{(1)}= M_{\Xi^{*}}-M_{\Omega}+\delta m_{10}^{(2)}.\\
\end{equation}
Here as $\delta m_{8}=\delta m_{10}^{(1)}=\delta m_{10}^{(2)}=0,$
eqs.(1) and (2) are the standard GOR. However,in the real world
\begin{equation}
\delta m_{8}=-26MeV,
\end{equation}
\begin{equation}
\delta m_{10}^{(1)}=4.5MeV,
\end{equation}
\begin{equation}
\delta m_{10}^{(2)}=9.6MeV,
\end{equation}
which stand for the deviations of the GOR from experimental data. The
motivation of our studies in this present paper is to calculate $\delta
m_{8}, \delta m_{10}^{(1)}$ and $\delta m_{10}^{(2)}$ in terms of the
Skyrme model$^{[4][5]} $ analytically.

   In the Skyrme model the baryon-octet and -decuplet emerge as
topological solitons (i.e.,skyrm-ions) in the SU(3)$\otimes$SU(3)
current algebra chiral Lagrangians. It is believed that this chiral
soliton model provides a reasonable dynamical mechanism for the
mass-splitting of SU(3)-baryons. In actual factor, through
calculations of the first order perturbation to the masses of baryons in
the SU(3) Skyrme model, one can re-establish the GOR$^{[5]}$. Thus it
could be expected that the high order corrections calculation in the
perturbations of the skyrmion quantum mechanics will show the values
of $\delta m_{8}, \delta m_{10}^{(1)}$ and $\delta m_{10}^{(2)}$.
In the present paper we will complete the calculations of the second
order corrections which should be the leading order to $\delta
m_{8}, \delta m_{10}^{(1)}$ and $\delta m_{10}^{(2)}$, and learn
some new physical implications from them.

   The facts that the perturbation calculations for (ud)- and (s)-flaver
symmetry breaking in both SU(3) group theory$^{[[1]}$ and skyrmion
theory$^{[5]}$ have shown the GOR successfuly imply that the
perturbation method seems to be reasonable and legitimate. However, it
should be pointed out that those successes are achieved under the
separation of the baryon-octet section and -decuplet section, and the
GOR are relative mass-relations. As one employed the perturbation
approach to calculate the absolute values of the baryon-masses, especially
the mass-difference between the octet and the decuplet in the Skyrme
model,  one would encounter
a serious difficulty that the pion decay
constant $F_{\pi}$ resulted from appropriate
adjusting the parameters in the model
is too small (almost 2/3 its experimental value)$^{[5]}$.
In the following we will
limit ourselve to deriving the relative mass-relations in the octet section
and the decuplet section respectivity by means of the perturbation approach.

   2. The Harmitonian of the SU(3) skyrmion quantum mechanics in the
SU(3) collective coordinate space is$^{[6]}$ (we use the notations in
ref.[6] hereafter)
\begin{eqnarray}
H &=& H_{0}+H^{\prime}\\
H_{0} &=& M_{s}+ \frac{1}{2b^{2}} ( \sum_{i=1}^{8}
L_{i}L_{i}-R_{8}^{2})+\frac{1}{2}(\frac{1}{a^{2}}-\frac{1}{b^{2}})
\sum_{A=1}^{3} R_{A}R_{A}+\frac{2 \delta}{\sqrt{3}} F_{\pi}R_{8}  \\
H^{\prime} &=& m (1-D_{8 8}^{(ad)}(A))
\end{eqnarray}
where $M_{s}$ is classical soliton's mass, $a^{2}$ and $b^{2}$ are the
soliton's moments of inertia, $ m, F_{\pi} $ and
$\delta$ are constants and parameters in the model, $D_{\mu
\nu}^{(ad)}$ denotes the regular adjoint representation functions of
SU(3), and $[L_{i},L_{j}]=if_{ijk}L_{k},[R_{i},R_{j}]=-if_{ijk}R_{k},
[L_{i},R_{j}]=0.$ In eq.(6) $H_{0}$ serves as the unperturbed
Hamiltonian, and $H^{\prime}$ as the perturbative part. It is easy to
see that $H_{0}$ is diagonal and the eigen-wavefunctions for $H_{0}$
are $^{[6]}$
\begin{equation}
|_{\mu \nu}^{\lambda} \rangle = (-1)^{s+s_{z}}\sqrt{\lambda}
D_{\mu,\nu}^{(\lambda)}(A)
\end{equation}
with
\begin{equation}
\mu =\left (\begin{array}{cc} I Y \\ I_{3}\end{array} \right ),
\nu =\left (\begin{array}{cc} S 1 \\ -S_{z}\end{array} \right ).
\end{equation}
$|_{\mu \nu}^{8} \rangle$ and $|_{\mu \nu}^{10} \rangle$ are
corresponding to the baryon-octet and -decuplet respectivily. As one
computes the matrix-elements of $\langle H^{\prime}\rangle$ a useful
formula is as following
\begin{equation}
\langle_{\mu_{2} \nu_{2}}^{\lambda_{2}}| D_{\mu
\nu}^{(\lambda )}(A)|_{\mu_{1} \nu_{1}}^{\lambda_{1}}
\rangle=(-1)^{s_{1}+s_{2}+s_{1z}+s_{2z}}
\sqrt{\frac{\lambda_{1}}{\lambda_{2}}} \sum_{\gamma} \left ( \begin{array}{ccc}
\lambda_{1} \lambda \lambda_{2\gamma} \\  \mu_{1} \mu \mu_{2}
\end{array} \right ) \left ( \begin{array}{ccc}
\lambda_{1} \lambda \lambda_{2\gamma} \\  \nu_{1} \nu \nu_{2}
\end{array} \right )
\end{equation}
with a standard notation for the Clebsch-Gordon coefficients. Here
$\gamma$ distingushes the independent irreducible representations
occurring in the reduction $(\lambda) \otimes (\lambda_{1})\rightarrow
(\lambda_{2}).$
   The mass of the baryon for $| k\rangle \equiv |_{\mu \nu}^{\lambda}
\rangle $ can be calculated in perturbation,
\begin{equation}
M_{k}=E_{k}^{(0)}+E_{k}^{(1)}+E_{k}^{(2)}+\cdots
\end{equation}
where
\begin{eqnarray}
E_{k}^{(0)} &=& \langle k| H_{0} |k \rangle, \\
E_{k}^{(1)} &=& \langle k| H^{\prime} |k \rangle, \\
E_{k}^{(2)} &=& \sum_{n\not= k} \frac{|\langle n| H^{\prime} |k \rangle
|^{2}}{ E_{k}^{(0)}-E_{n}^{(0)}}.
\end{eqnarray}
 $E_{k}^{(0)}$ and $E_{k}^{(1)}$ have been known in
literatures$^{[6]}$. Our object is to calculate $E_{k}^{(2)}$ which is
related to the non-diagonal matrix elements of $D_{8,8}^{(8)}(A)$
between the baryon-states (to see eq.(15)). We will study the cases of
baryon-octet and -decuplet respectivety.

   3. The baryon-octet section: Noticing SU(3)-multiplets decumposition
formula
\begin{equation}
8 \otimes 8=27 \oplus 10 \oplus 10^{*} \oplus 8_{F} \oplus 8_{D}\oplus
1
\end{equation}
we have (see eq.(15))
\begin{equation}
|n \rangle \in \{ 27, 10, 10^{*}, 1 \}.
\end{equation}
As $|k\rangle \in$ nucleon, the nonzero $(n\not= k)$ $\langle n|D_{8,8}^{(8)}
|k\rangle$-matrix elements are as following
\begin{eqnarray}
\langle D_{\mu \nu}^{(10^*)}|D^{(8)}_{
 \left( \begin{array}{c} 00 \\ 0
\end{array} \right)
 \left( \begin{array}{cc} 0 0 \\ 0
\end{array} \right)}
|D_{\left( \begin{array}{cc} \frac {1}{2} 1 \\ \frac{1}{2}
\end{array} \right)\left( \begin{array}{cc} \frac{1}{2} 1 \\ \frac{1}{2}
\end{array} \right)}^{(8)} \rangle &=& \sqrt{\frac {1}{20}}, \\
\langle D_{\mu \nu}^{(27)}|D_{\left( \begin{array}{cc} 0 0 \\ 0
\end{array} \right)\left( \begin{array}{cc} 0 0 \\ 0
\end{array} \right)}^{(8)}|D_{\left( \begin{array}{cc} \frac{1}{2} 1 \\
 \frac{1}{2}
\end{array} \right)\left( \begin{array}{cc} \frac{1}{2} 1 \\
 \frac{1}{2}
\end{array} \right)}^{(8)} \rangle &=& \sqrt{\frac {3}{50}},
\end{eqnarray}
here eq.(11) and the SU(3) CG coefficients listed in ref.[7] have been
used. Since the spins of $|k \rangle $ and $|n \rangle $ are same, from
eqs.(7) and (13) we have
\begin{equation}
E_{k}^{(0)}-E_{n}^{(0)}=\frac{1}{2b^{2}}(C_{2}(k)-C_{2}(n)),
\end{equation}
where $C_{2}(k)$ denotes the Casimir operator for the $k$-dimensional
irreducible representation of SU(3). Noticing $C_{2}(8)=3, C_{2}(10^*)=6$
and $C_{2}(27)=8,$ then
\begin{eqnarray}
E_{8}^{(0)}-E_{10^*}^{(0)} &=&-\frac{3}{2b^{2}} \\
E_{8}^{(0)}-E_{27}^{(0)} &=&-\frac{5}{2b^{2}}
\end{eqnarray}
Combining above with the known results in refs.[5][6], we have the
nucleon's mass to the second order in perturbation,
\begin{equation}
M_{N}=M_{8}-\frac{3}{10} m-\frac{43}{750} g
\end{equation}
where
\begin{eqnarray}
M_{8}&=&\langle H_{0} \rangle _{\lambda =8}, \\
g &=& m^{2} b^{2}.
\end{eqnarray}
Through the similar computations, we can obtain the masses of
$\Lambda,\Sigma$ and $\Xi$ to the second order in perturbation,
\begin{eqnarray}
M_{\Lambda}&=&M_{8}-\frac{1}{10}m-\frac{9}{250}g  \\
M_{\Sigma}&=&M_{8}+\frac{1}{10}m-\frac{37}{750}g  \\
M_{\Xi}&=&M_{8}+\frac{1}{5}m-\frac{3}{125}g
\end{eqnarray}
where $M_{8},m$ and $g$ are model-dependent variables. Eliminating
$M_{8},m$ and $g$  in eqs.(23),(26),(27) and (28), we have
\begin{equation}
2(M_{N}+M_{\Xi})=3M_{\Lambda}+M_{\Sigma}+\frac{2}{13}
(M_{N}+M_{\Sigma}-2M_{\Lambda})
\end{equation}
Comparing eq.(29) with eq.(1), we can see that the last term in RHD of
eq.(29) serves as a correction coming from the skyrmion dynamics to GOR
in octet section. Thus we get the desired result
\begin{equation}
\delta m_{8}=\frac{2}{13}(M_{N}+M_{\Sigma}-2M_{\Lambda})
\end{equation}
By use of the mass relation of eq.(29), the mass-variables $(M_{N},
M_{\Sigma}, M_{\Lambda})$ in eq.(30) can be changed to be $(M_{\Xi},
M_{\Sigma}, M_{\Lambda})$, or $(M_{N},
M_{\Sigma}, M_{\Xi})$, or $(M_{N},
M_{\Xi}, M_{\Lambda})$. Namely, we can also write $\delta m_{8}$ as
followings,
\begin{eqnarray}
\delta m_{8}&=& \frac{1}{12}(3M_{\Sigma}-2M_{\Xi}-M_{\Lambda}), \\
             &=& \frac{2}{35}(5M_{\Sigma}-M_{N}-4M_{\Xi}), \\
             &=& \frac{1}{15}(6M_{N}+4M_{\Xi}-10M_{\Lambda}).
\end{eqnarray}
These four expressions of $\delta m_{8}$ are equivalent each other in
principle. However, since eq.(29) is still not an exact identity of
masses when experimental mass-data are inputs, the numerical results
given by eqs.(30)-(33) will be not exactly same, i.e.,there is a small
error for the theoretical prediction of $\delta m_{8}$. From
eqs.(30)-(33) we get
\begin{equation}
\delta m_{8} =-15.35 \pm 1.32 MeV.
\end{equation}
This second order correction to $\delta m_{8}$ agrees with the
experiment qualitatively, and in quantitative respect it is
compatible with the data showing in eq.(3) roughly.

     4. The decuplet section. As $|k \rangle$ (in eq.(15)) belongs to
decuplet, from the SU(3)-decomposition formula
\begin{equation}
10 \otimes 8 = 35 \oplus 27 \oplus 10 \oplus 8
\end{equation}
we have
\begin{equation}
|n \rangle \in \{ 35, 27, 8 \}.
\end{equation}
Using again the formula eqs.(11),(20),CG coefficients of SU(3)$^{[7]}$and
$C_{2}(35)
=12$,
we get the masses of the baryon-decuplet to second order in perturbation,
\begin{eqnarray}
M_{\Delta} &=& M_{10}-\frac{1}{8} m-\frac{85}{672} g,  \\
M_{\Sigma^{*}} &=& M_{10}-\frac{26}{336} g,  \\
M_{\Xi^{*}} &=& M_{10}+\frac{1}{8} m-\frac{9}{224} g,  \\
M_{\Omega} &=& M_{10}+\frac{1}{4} m-\frac{5}{336} g,
\end{eqnarray}
where
\begin{equation}
M_{10}=\langle H_{0} \rangle _{\lambda =10} .
\end{equation}
Here, there are four equations and three unkowns. By solving these
over-determinant equations, a constrained equation is left, which is
\begin{equation}
M_{\Omega}-M_{\Delta}=3(M_{\Xi^{*}}-M_{\Sigma^{*}})
\end{equation}
Eq.(42) is called Okubo relationship which holds to second order in
flavor breaking as shown by Okubo long ago$^{[8]}$ and by Morpurgo
recently$^{[2]}.$ Here, we reach such conclusion again in the skyrmion
formlism. The eqs.(37)-(40) show that the equal spacing rule
for the decuplet (i.e.,GOR in decuplet section) no longer holds. Using
the definitions of $\delta m_{10}^{(1)}$ and $\delta m_{10}^{(2)}$
(eq.(2)), and the mass-splitting formulas eqs.(37)-(40), we have
$\delta m_{10}^{(1)} =\frac{1}{84}g $, $\delta m_{10}^{(2)}=\frac{1}{42}
 g $. Then we get
\begin{equation}
\frac{\delta m_{10}^{(2)}}{\delta m_{10}^{(1)}} =2,
\end{equation}
which is quite in agreement with the experiment (to see eqs.(4)(5))
\begin{equation}
\left( \frac{\delta m_{10}^{(2)}}{\delta m_{10}^{(1)}} \right)_{expt}
=2.1.
\end{equation}
 From eqs.(37)-(40), $g$ can be found out to be
$84(2M_{\Sigma^{*}}-M_{\Xi^{*}}-M_{\Delta}).$ Then we get
\begin{equation}
\delta m_{10}^{(1)}= 2M_{\Sigma^{*}}-M_{\Xi^{*}}-M_{\Delta}.
\end{equation}
By using Okubo relationship eq.(42) the mass-variables in eq.(45) could
be changed. The other three expressions for $\delta m_{10}^{(1)}$ read
\begin{eqnarray}
\delta m_{10}^{(1)} &=&\frac{1}{3}(3M_{\Xi^{*}}-2M_{\Omega}-M_{\Delta}),
   \\
   &=&\frac{1}{3}(3M_{\Sigma^{*}}-2M_{\Delta}-M_{\Omega}),  \\
   &=&2M_{\Xi^{*}}-M_{\Sigma}-M_{\Omega}.
\end{eqnarray}
These four $\delta m_{10}^{(1)}$-expressions are equivalent in
principle, but their numerical results are not exactly same since
eq.(42) is approximate just like the octet case above. Thus we have
\begin{equation}
\delta m_{10}^{(1)}=7.05\pm 2.55 MeV
\end{equation}
and
\begin{equation}
\delta m_{10}^{(2)}=2\delta m_{10}^{(1)}=14.1\pm 5.1 MeV,
\end{equation}
where eq.(43) is used. They agree with the experimental data
shown in eqs.(4) and (5).

    5. The picture that the baryons are regarded as the chiral solitons
has extensively been investegated during last ten years$^{[9]}$. It is
well known that the quantum mechanics of SU(3)-skyrmion is basic
formlism to deal with (ud)- and (s)-flavor breaking in the model.
Following the investigations of the first order in flavor breaking in
this quantum mechanics problem, we have completed the calculations of
the second order corrections to the baryon-mass splitting and shown
that the results support the soliton pictures. Especially, since high
order in perturbations of quantum mechanics are intimately related to
the $H^{\prime}$-matrix elements between various eigenstates of $H_{0}$
(see eqs.(12)-(15)), our above results can be regarded as signals of
the existence of some obscure SU(3)-multiplets: 27-let (with spin
$\frac{1}{2}$ or $\frac{3}{2}$), 10$^*$-let (with spin $\frac{1}{2}$) and
35-let (with spin $\frac{3}{2}$). It is easy to be sure that the wave
functions of these obscure SU(3)-states satisfy the constrained condition
coming
from the Wess-Zumino term in the QCD effective Lagrangian, i.e.,the
spin-hypercharge $Y_{R}=1 ^{[5]}$. So they should be physical states
in QCD. However, there is not yet any experimental evidence showing them as
particle-like resonances so far. A possible explanation for it seems to
be that the width of the corresponding resonances is so wide that the
identifications for them become very difficult in experiments. Indeed,
the non-zero $H^{\prime}$-matrix elements related to these obscure
SU(3)-states (they imply some new strong decay mode)
will increase the widths of
the resonances and make them much wider than usual decuplet particle's
(note: $\langle octet|H^{\prime}|decuplet \rangle =0 )$. But, just
these non-zero $H^{\prime}$-matrix elements can bring contributions to $
\delta m_{8}, \delta m_{10}^{(1)}$ and $ \delta m_{10}^{(2)}$ through
the second order perturbations. This may be the reason why the obscure
SU(3)-multiplets can be seen through the corrections of $\delta m_{8},
\delta m_{10}^{(1)}$ and $ \delta m_{10}^{(2)}$, but not seen in
particle-like resonances directly.As a puzzle,it was attended long ago
whether there exist some SU(3) baryon-multiplets besides the ordinary
octet and decuplet in the flavor SU(3) theory.The above positive conclusion
on it could be interesting.

   For the Skyrme model, the quark-variable freedoms have been
integrated out. If the quark-descriptions on the SU(3) skyrmions are
resumed, then all of obscure SU(3) multiplets mentioned above should
belong to six-quark states. Thus we could see that even numbers of
quarks could form a baryon with half-integer spin. This would be a new
feature intimately related to the Wess-Zumino term in the skyrmion
physics.

  Finally,we would like to remark that in the present paper
 the mass spectrums of the octet section and
the decuplet section are calculated in perturbation respectively
and independently
.However,due to the limitations of the perturbation method for the
(ud)- and (s)-flavor symmetry breaking,it can not be extended to the case
of the octet and the decuplet combined.
If one wants to get a complete mass spectrum (including
octet and decuplet simultaneously ),the skyrmion quantum mechanics
problem in the SU(3) collective coordinate space ( see eqs.(6)--(8))
have to be solved exactly.Unfortunately,this goal can merely be reached
numerically$^{[10]}$,instead of analytically.Therefore,in order to reveal
the physical implications explicitly,our analytical analyses
in the above seems necessary.

\vskip3.5cm


